\documentclass[aps,prl,showpacs,twocolumn]{revtex4}
\usepackage{bm, graphicx, amsmath}
\usepackage{bbm}
\usepackage{amssymb,amsmath,amsthm}
\usepackage[mathscr]{eucal}
\usepackage{graphicx}

\newcommand{\id}{{\mathbb I}}
\newcommand{\tr}{{\rm tr}\,}
\newcommand{\ket}[1]{\left|{#1}\right\rangle}

\newcommand{\Qb}{\bar Q}
\theoremstyle{definition}

\theoremstyle{remark}

\begin{document}

\title
{Optimal discrimination of quantum states with a fixed rate of inconclusive outcomes}
\author{E. Bagan$^{1,2}$, R. Mu\~{n}oz-Tapia$^{1}$, G. A. Olivares-Renter\'{i}a$^{2}$, and J. A. Bergou$^{2}$}
\affiliation{$^{1}$F\'{i}sica Te\`{o}rica: Informaci\'{o} i Fen\`{o}mens Qu\`antics, Universitat Aut\`{o}noma de Barcelona, 08193 Bellaterra (Barcelona), Spain \\ $^{2}$Department of Physics and Astronomy, Hunter College of the City University of New York, 695 Park Avenue, New York, NY 10065, USA}

\begin{abstract}
In this paper we present the solution to the problem of optimally discriminating among quantum states, i.e., identifying the states with maximum probability of success when a certain fixed rate of inconclusive answers is allowed. By varying the inconclusive rate, the scheme optimally interpolates between Unambiguous and Minimum Error discrimination, the two standard approaches to quantum state discrimination. We introduce a very general method that enables us to obtain the solution in a wide range of cases and give a complete characterization of the minimum discrimination error as a function of the rate of inconclusive answers.  A critical value of this rate is identified that coincides with the minimum failure probability in the cases where unambiguous discrimination is possible and provides a natural generalization of it when states cannot be unambiguously discriminated. The method is illustrated on two explicit examples: discrimination of two pure states with arbitrary prior probabilities and discrimination of trine states.

\end{abstract}


\maketitle

State discrimination has long been recognized to play a central role in quantum information and quantum computing. In these fields the information is encoded in the state of quantum systems,
thus, one often needs to identify in which of~$N$ known states~$\{\rho_{i}\}_{i=1}^N$ one such system was prepared. If the possible states are mutually orthogonal this is an easy task: we just set up detectors along these orthogonal directions and determine which one clicks (assuming perfect detectors). However, if the states are not mutually orthogonal the problem is highly nontrivial and optimization with respect to some reasonable criteria leads to complex strategies often involving generalized measurements. Finding such optimal strategies is the subject of state discrimination.

%
%
The two fundamental state discrimination strategies are discrimination with minimum-error (ME) and unambiguous discrimination (UD). In ME, every time a system is given and a measurement is performed on it a conclusion must be drawn about its state. Accordingly, the measurement is described by a $N$-element positive operator valued measure (POVM) $\Pi=\{\Pi_i\}_{i=1}^N$, where each element represents a conclusive outcome. Errors are permitted and in the optimal strategy the probability of making an error is minimized \cite{helstrom}. In UD, no errors are tolerated but at the expense of permitting an inconclusive measurement outcome, represented by the positive operator~$\Pi_0$. Hence, the corresponding POVM is $\Pi=\{\Pi_i\}_{i=0}^{N}$. When $\Pi_0$ clicks we do not learn anything about the state of the system and in the optimal strategy the probability of the inconclusive outcome is minimized~\cite{unambiguous}. It has been recognized that states can be discriminated unambiguously only if they are linearly independent \cite{chefles1}. Discrimination with maximum confidence (MC) can be applied to states that are not necessarily independent and for linearly independent states it reduces to unambiguous discrimination \cite{croke,herzog1}, so the MC scheme can be regarded as a generalized UD strategy.


It is clear that UD (or MC for linearly dependent states) and ME are the two extremes of a more general scheme that can be approached by relaxing the conditions at either end.  That is, we may reduce the optimal error rate in the ME approach by allowing  certain fixed rate~$Q$ of inconclusive outcomes \cite{chefles2,zhang,fiurasek,eldar},  or, starting from~UD,  allow for some fixed rate $P_{e}$ of errors to occur~\cite{touzel,hayashi,sugimoto} and hence reduce~$Q$. The first approach yields the minimal error rate as a function of the given inconclusive rate~$P^{\rm min}_{e}(Q)$, while the second yields the minimum $Q$ as a function of the given error rate $Q^{\rm min}(P_{e})$. Since the two expressions are the result of optimizations with different conditions, 
it is not immediately obvious, yet true, that one is the inverse of the other~\cite{herzog2}.  Notice that these general scenarios encompass many practical situations, where resources are scarce and one can only afford a limited rate of inconclusive outcomes, or where the error rate must be kept below certain level but need not be strictly vanishing.
 
The function $P^{\rm min}_{e}(Q)$ has never been solved in full generality. Analytical solutions have been provided for two pure states with equal occurrence probabilities, otherwise only useful bounds were established \cite{chefles2,zhang, fiurasek,eldar}. The second approach has recently been solved for the case of two pure states with arbitrary occurrence probabilities~\cite{sugimoto}, but it is not obvious whether the method can be extended to other cases involving, e.g., mixed states, more than two states, etc. Furthermore the connection with the first approach remained unnoticed, and so remains the connection with the MC scheme.

Here we state some fundamental features of the Fixed Rate of Inconclusive Outcome (FRIO) scheme, in particular of the function $P^{\rm min}_{e}(Q)$, and show its links to the other schemes discussed above. We thus give a unified and quite complete picture of quantum state discrimination. 
Perhaps most important, we also present a very general method  to obtain $P^{\rm min}_{e}(Q)$.  This in turn provides also the solution,~$Q^{\rm min}(P_e)$, to the second approach.

The idea is to first introduce a suitable transformation which formally eliminates the inconclusive part of the problem and reduces it to a ME problem. Since the optimal solution to the ME problem is known for many special cases, we can immediately adopt it as our starting point. This formal solution still depends on free parameters of the transformation. In the next step we carry out a second optimization with respect to these free parameters which then yields the complete solution.

For clarity's sake, we will introduce FRIO discrimination on the case of two states, $\rho_1$ and $\rho_2$, with general \textit{a priori} probabilities  $\eta_{1}$ and $\eta_{2}$ such that \mbox{$\eta_{1} + \eta_{2} = 1$.} The generalization to more than two states is straightforward.
We~introduce a three element POVM $\Pi=\{\Pi_0,\Pi_1,\Pi_2\}$, with $\Pi_1+\Pi_2+\Pi_0=\id$, where $\Pi_{1(2)}$ identifies $\rho_{1(2)}$, while~$\Pi_{0}$ corresponds to the inconclusive outcome. The average success, error and inconclusive probabilities are:
\begin{eqnarray}
P_s&=&\tr(\eta_1\rho_1 \Pi_1)+\tr(\eta_2\rho_2 \Pi_2) ,\\
P_e&=&\tr(\eta_1\rho_1 \Pi_2)+\tr(\eta_2\rho_2 \Pi_1),\\
Q&=&
\tr(\rho\Pi_0),
\label{Q}
\end{eqnarray}
where $\rho \equiv \eta_1\rho_1+\eta_2\rho_2$,
and we assume a fixed $Q$. Clearly we have $P_s+P_e+Q=1$. 
The optimal strategy minimizes $P_{e}$ under the constraint that $Q$ is fixed, yielding $P^{\rm min}_e(Q)$.
%
%

It follows from the above definitions that $P^{\rm min}_e(Q)$ is a convex function. To show this, let us assume that $\Pi'$ ($\Pi''$) is the optimal POVM for a rate $Q'$ ($Q''$) of  inconclusive outcomes. Then, the mixed strategy that consists in performing the measurement $\Pi'$ ($\Pi''$) with probability~$p$ ($\bar p=1-p$), i.e., the measurement strategy given by the POVM~$p\Pi'+\bar p\, \Pi''$, has error probability~$p P^{\rm min}_e(Q')+\bar p P^{\rm min}_e(Q'')$ for a rate~$Q=p\,Q'+\bar p\,Q''$ of inconclusive outcomes. Since~$p\Pi+\bar p\, \Pi'$ is not necessarily optimal for $Q$, we have the convexity inequality,
%
\begin{equation}
P^{\rm min}_e(Q=p\,Q'+\bar p\, Q'')\le p P^{\rm min}_e(Q')+\bar p P^{\rm min}_e(Q'').
\label{conv ineq}
\end{equation}

Convexity implies the following useful properties~\cite{convex}: 
i) $P^{\rm min}(Q)$ is continuous in~$(0,1]$; 
ii) $P^{\rm min}(Q)$ is differentiable in~$(0,1)$ except, at most, at countably many points;  
iii) Left and right derivatives exist, they satisfy $(P^{\rm min}_e)'(Q_-)\le (P^{\rm min}_e)'(Q_+)$, and are monotonically non-decreasing functions of $Q$; 
iv) In~addition, since~$P^{\rm min}_e(Q)\ge0$ and $P^{\rm min}_e(1)=0$, we have that~$(P^{\rm min}_e)'(Q_{\pm})\le 0$ and~$P^{\rm min}_e(Q)$ is non-increasing. 

Because of these properties, there must exist a critical rate $Q_{\rm c}$ such that for $Q_{\rm c}\le Q\le1$ the right derivative~$(P^{\rm min}_e)'(Q_+)$ is constant and takes its maximum value $-\alpha\equiv (P^{\rm min}_e)'(Q_{{\rm c}+})$. Hence, 
\begin{equation}
P^{\rm min}_e(Q)=\alpha(1-Q),\qquad  Q_{\rm c}\le Q\le1 .
\label{alpha}
\end{equation}
Note that $P^{\rm min}_e(Q)/(1-Q)$ is the error probability conditioned on obtaining a conclusive answer. Eq.~(\ref{alpha}) states that this conditioned probability becomes a constant for inconclusive rates larger than $Q_{\rm c}$.  The quantity $\bar\alpha\equiv1-\alpha$ can be interpreted as a confidence~$C$. Indeed, for symmetric states, such as those in our second example below,~$\bar\alpha$ is the maximum confidence and $Q_{\rm c}$ coincides with the minimum rate of inconclusive outcome in the MC scheme: $Q_{\rm c}=Q^{\rm MC}\kern-.2em$. Other links with the MC scheme will be given in~\cite{bagan2}. Note that when the states can be unambiguously discriminated, as in the case of two pure sates, we have~$\alpha=0$ ($C=\bar\alpha=1$); $Q_{\rm c}$ is equal to the optimal inconclusive (or failure) probability in the UD scheme:~$Q_{\rm c}=Q^{\rm UD}$; and $P^{\rm min}_e(Q)=0$ for~$Q_{\rm c}\le Q\le1$.

In order to obtain optimal FRIO measurement strategies, it suffices to find  optimal POVMs in the region of~$Q$ where $P^{\rm min}_e(Q)$ is strictly convex. Outside this region, the proof given above shows that the best measurements will be trivial convex combinations of those optimal POVMs,  i.e., mixed strategies. This provides an important simplification to our optimization problem (see examples below) through the following theorem: {\em At values of $Q$ such that~$P^{\rm min}_e(Q)$ is strictly convex (i.e.,~where optimal strategies are pure), the element~$\Pi_0$ of the optimal POVM  necessarily has a zero eigenvalue}. 

%
%
%
This is so because if all the eigenvalues of~$\Pi_0$ were non-zero, it could be written as $\Pi_0=\bar p\, \Pi'_0+p\id$ for some values of~$p$, $0<p<1$, and some non-negative operator $\Pi'_0$. Note that $\Pi=\bar p\,\Pi'+p\Pi''$, where $\Pi'=\{\Pi'_0,{\bar p\,}^{-1}\Pi_{1(2)}\}$ and~$\Pi''=\{\id,0,0\}$ are proper POVMs (the latter is the trivial strategy with no conclusive outcomes). Thus, $\Pi$ would define a mixed, rather than a pure, strategy and~$P^{\rm min}_e(Q)$ would not be strictly convex. 

We next show how to transform the problem of finding the optimal FRIO strategy to a minimum error problem with no inconclusive outcome. The starting point is to write
$
\Pi_1+\Pi_2=\id-\Pi_0 \equiv \Omega.
$
Multiplying both sides of this equation by 
$\Omega^{-1/2}$  we have%
~$\tilde\Pi_1+\tilde\Pi_2=\id$,
where
\begin{equation}
\tilde\Pi_i=\Omega^{-1/2}\Pi_i \,\Omega^{-1/2}\ge0 .
\label{tildeops}
\end{equation}
So $\{\tilde\Pi_1,\tilde\Pi_2\}$ is a POVM.
Note that $\Omega^{-1/2}$ exists unless~$\Pi_0$ has a unit eigenvalue, in which case we address the problem differently (see later).
Defining the normalized transformed states $\tilde\rho_i$ and  \emph{a priori} probabilities $\tilde\eta_i$ as
\begin{equation}
\tilde\rho_i={\Omega^{1/2}\rho_i\Omega^{1/2}\over \tr(\Omega\rho_i)} ,  \qquad  \tilde\eta_i={\eta_i\tr(\Omega\rho_i)\over\Qb},
\label{tildestates}
\end{equation}
where $\Qb\equiv1-Q$,
the success and error probabilities read~$P_{s(e)}=\bar Q \tilde P_{s(e)}$, where 
\begin{eqnarray}
\tilde{P}_{e} &=& \tr(\tilde\eta_1\tilde\rho_1 \tilde\Pi_2)+\tr(\tilde\eta_2\tilde\rho_2 \tilde\Pi_1) ,
\label{tildePe}
\end{eqnarray}
and $\tilde{P}_{s} = 1 - \tilde{P}_{e}$. 
Equation \eqref{tildePe} and $\tilde\Pi_1+\tilde\Pi_2=\id$ define a ME discrimination problem for the transformed states and \emph{priors} given in Eq.~\eqref{tildestates}. 
Further minimization over the choice of~$\Pi_0$, such that $\tr\rho\Pi_0=Q$, gives the desired result
%
$P^{\rm min}_e(Q)=\min \Qb\,\tilde P^{\rm ME} _e$. 
%
The optimal solution to this ME discrimination problem is well known~\cite{helstrom}: $\tilde P^{\rm ME}_e=(1-\parallel\tilde{\eta}_{2} \tilde{\rho}_{2} - \tilde{\eta}_{1} \tilde{\rho}_{1} \parallel_1)/2 $, where~$\parallel\kern-.2em\cdot\kern-.2em\parallel_1$ is the trace norm. 
%
An analogous formula does not exist for more than two states. 
%
%
However, explicit solutions to the ME discrimination of $N$ states~$\{\tilde\rho_i\}_{i=1}^N$ are known in some particular cases, e.g., symmetric states. Then one can carry out the last minimization over~$\Pi_0$.

We next illustrate the method on the example of two pure states, $\rho_{1} = |\psi_{1}\rangle\langle\psi_{1}|$ and $\rho_{2} = |\psi_{2}\rangle\langle\psi_{2}|$ with general \textit{a priori} probabilities  $\eta_{1}$ and $\eta_{2}$. 
Since $\tilde\rho_i=|\tilde\psi_i\rangle\langle\tilde\psi_i|$ are also pure states, 
we can use the more explicit expression
\begin{equation}
\tilde{P}^{\rm ME}_{e}={1\over2}\left(1-\sqrt{1-4\tilde\eta_1\tilde\eta_2|\langle\tilde\psi_1|\tilde\psi_2\rangle|^2}\;\right).
\label{tildePe2}
\end{equation}
%
%
Since two pure states can be unambiguously discriminated, the critical probability $Q_{\rm c}$ coincides with the optimal failure probability~$Q^{\rm UD}$
which reads~\cite{bergourev}
\begin{equation}
    \label{Qmax}
    Q_{\rm c} = \left\{ \begin{array}{l}
    \eta_{1}+\eta_{2}\cos^{2}\theta, \mbox{if $\displaystyle \eta_{1} <
    \frac{\cos^{2}\theta}{1+\cos^{2}\theta}\equiv\eta_1^{(l)} $,}  \\ [.7em]
    \eta_{2}+\eta_{1}\cos^{2}\theta, \mbox{if $\eta_1>\displaystyle \frac{1}{1+\cos^{2}\theta}
    \equiv\eta_{1}^{(r)}$,}\\[.7em] 
   2\sqrt{\eta_{1}\eta_{2}}\cos\theta\equiv Q_0, \mbox{if $\eta_1^{(l)}\le\eta_1\le\eta_1^{(r)}$}\ ,
    \end{array}
    \right. 
\end{equation}
where $|\langle\psi_{1}|\psi_{2}\rangle| \equiv \cos \theta$ is the overlap of the states.
%
  
We now set out to find the optimal pure strategy, i.e., the optimal POVM that will minimize $P_{e}$ for a fixed $Q$ in the interval $0 \leq Q \leq Q_{c}$.
As the Hilbert space spanned by two pure states is two-dimensional and the optimal~$\Pi_{0}$ has a zero eigenvalue, it is effectively a positive rank one operator. We denote the positive eigenvalue by $\xi$, $0\leq \xi \leq 1$, the eigenstate belonging to $\xi$ by $|0\rangle$ and the orthogonal state by $|1\rangle$. In this basis $\Pi_0=\xi|0\rangle\langle0|$, and~$\Omega=\bar\xi|0\rangle\langle0|+|1\rangle\langle1|$, where $\bar\xi\equiv1-\xi$.
If we also write the input states in this basis,~%
\mbox{$
|\psi_i\rangle=c_i|0\rangle+s_i|1\rangle
$},
where $c_i\equiv\cos\theta_i$, $s_i\equiv\sin\theta_i$, the transformed states and priors can be trivially obtained from Eq.~\eqref{tildestates}, and after obvious simplifications we can write
%
%
\begin{equation}
\Qb\,\tilde{P}^{\rm ME}_e={1\over2}\left\{\Qb-\sqrt{\Qb^2-4\eta_1\eta_2(\cos\theta-\xi c_1c_2)^2} \right\} ,
\label{tildePe3}
\end{equation}
%
where we used that $\theta_{1} - \theta_{2} \equiv \theta$. It follows from~(\ref{Q}) that
\begin{equation}
\label{xi}
\xi=\frac{Q}{\eta_1c_1^2+\eta_2 c_2^2}.
\end{equation}
Hence Eq.~\eqref{tildePe3} depends only on one parameter, say~$\theta_1$, which determines the orientation of $\Pi_0$ relative
to that of the two pure states. The minimization~over~$\Pi_0$ (equivalently, over $\theta_1$) simplifies considerably using Eq.~(\ref{xi}) and defining
 %
$ c_1\,\eta_1^{1/2}(\eta_1c_1^2+\eta_2c_2^2)^{-1/2}\equiv\cos\varphi$, and~$c_2\,\eta_2^{1/2}(\eta_1c_1^2+\eta_2c_2^2)^{-1/2}\equiv\sin\varphi$.
%
 %
%
%
%
%
%
The resulting expression is minimum for~$\varphi=\pi/4$, yielding
\begin{equation}
P^{\rm min/max}_{e/s}\!=\!{1\over2}\!\left\{\Qb\mp \sqrt{\Qb{}^2 - \left(Q_{0} - Q\right)^2}\right\} ,\; Q\le Q_0,
\label{PeSol1}
\end{equation}
where $Q_0$ was introduced in the third line of~\eqref{Qmax}.
This is the optimal error/success rate for an intermediate range of the prior probabilities. 
%
One can invert $P^{\rm min}_e(Q)$ to obtain $Q$ and, in turn, $P^{\rm max}_s$ as function of~$P_e$, $P^{\rm max}_s(P_e) = [P_e^{1/2}+(1-Q_0)^{1/2}]^2$, in agreement with~\cite{hayashi}.
%


For the validity of these results,~$\xi\le1$ must hold. This condition determines the range of  \emph{priors} for which Eq.~\eqref{PeSol1} is valid.
The definitions of $\cos\varphi$ and $\sin\varphi$ after Eq.~(\ref{xi}) give
${\sqrt\eta_2 c_2=\sqrt\eta_1 c_1}$ for $\varphi=\pi/4$. 
Some straightforward algebra leads to 
$\eta_1c_1^2=\eta_2c_2^2=\eta_1\eta_2\,{\sin^2\theta/( 1-Q_0)}$,
and~Eq.~\eqref{xi} yields
$
\xi=(1-Q_0)Q/(2\eta_1\eta_2\sin^2\theta) 
$.
Setting $\xi = 1$ defines a threshold, 
\begin{equation}
Q_{\rm th}\equiv {2\eta_1\eta_2\sin^2\theta/(1-Q_0)}.
\label{Qth}
\end{equation}
Hence~$\xi \leq 1$ if $Q \leq Q_{\rm th}$ and $\xi = 1$ if $Q > Q_{\rm th}$. 

\begin{figure}[ht,floatfix]
    \centering
      \includegraphics[height=4.5 cm]{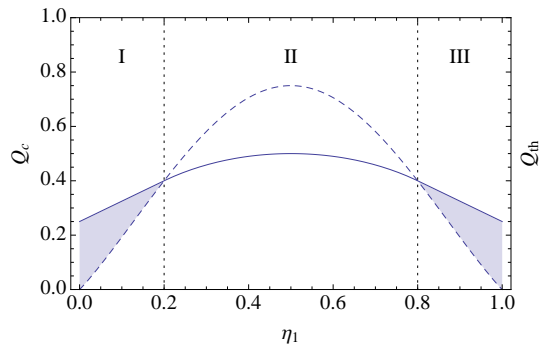}    
     \caption{$Q_{\rm c}$ (solid line) and $Q_{\rm th}$ (dashed line) vs.~$\eta_{1}$ for $\theta=\pi/3$. The area of interest lies under the solid line. The regions~I,~II and~III are defined in Eq.~(\ref{Qmax}).}
     \label{Fig1}
     \end{figure}

In Fig.\,\ref{Fig1} we plot $Q_{\rm c}$ and~$Q_{\rm th}$ vs.~$\eta_{1}$ together for a fixed overlap, $\cos\theta = 0.5$ ($\theta = \pi/3$).
The two curves intersect at 
$
\eta_1=\eta_{1}^{(l)}
$
and
$
\eta_1=\eta_{1}^{(r)}
$,
the same points as in Eq.~\eqref{Qmax}.
The interval $0 \leq \eta_{1} \leq 1$ is thus divided  into three regions. In regions~I and~III, we have~\mbox{$Q_{\rm th} < Q_{\rm c}$} and the solution~\eqref{PeSol1} is valid for $0 \leq Q < Q_{\rm th}$ only. In Region II, $\eta_{1}^{(l)} \leq \eta_{1} \leq \eta_{1}^{(r)}$, we have~\mbox{$Q_{\rm c} = Q_{0} < Q_{\rm th}$} and the solution \eqref{PeSol1} is valid for the entire $0 \leq Q \leq Q_{\rm c}$ range. 

In the shaded parts of regions I and II one has $Q_{\rm th} \leq Q \leq Q_{\rm c}$ and, necessarily, $\xi=1$. 
Hence,~$\Pi_0=|0\rangle\langle0|$ and $\Omega=|1\rangle\langle1|$ are projectors. Therefore,~$\Omega^{-1/2}$ does not exist in these areas and the case needs special consideration. 

%
%
%
The calculation of the error probability is most easily performed by realizing that $\Pi_1$ and $\Pi_2$ become degenerate, both must be
proportional to $\Pi_{d}=|1\rangle\langle1|$. The three-element POVM becomes a standard two element projective measurement,
$
\{\Pi_d=|1\rangle\langle1|
,
 \Pi_0=|0\rangle\langle0|
 \}
$.
We identify a click in~$\Pi_{d}$ with $\rho_{1}$ ($\rho_{2}$)  if $\eta_1\ge\eta_2$ ($\eta_2\ge\eta_1$), so $P_{e(s)}=\eta_2 s_2^2, \quad P_{s(e)}=\eta_1 s_1^2$, with $Q=1-P_e-P_s$. These equations completely determine the solution. There is nothing to optimize here, so we drop the superscript $\rm min$ in what follows. $\theta_1-\theta_2=\theta$ immediately gives $Q(P_e)$ as
\begin{equation}
Q\!=\!1\! -\! P_{e}\! - \eta_{1(2)}\!\!\left(\!\sqrt{\!{P_e\over \eta_{2(1)}}}\cos\theta\!\pm\!\sqrt{1\!-\!{P_e\over\eta_{2(1)}}}\,\sin\theta\!\right)^{\!2}\!.
\label{PeSol2}
\end{equation}
Inverting this equation gives $P_e(Q)$  but the resulting expression is not particularly insightful and we will not give it here. 
Note that for $P_e=0$ (UD limit) one has~$Q=Q_{\rm c}$, given by the first and second lines in Eq.~(\ref{Qmax}), and for $Q = Q_{\rm th}$, $P_{e}$ reduces to \eqref{PeSol1}, as it should.
%
   

%

\begin{figure}[ht,floatfix]
    \centering
      \includegraphics[height=4.5 cm]{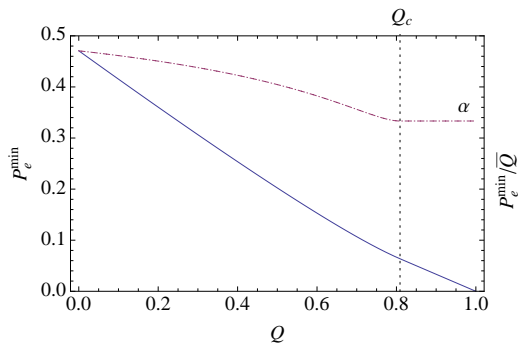}    
     \caption{Trine-state minimum error $P^{\rm min}_e(Q)$ (solid line) and minimum error conditioned on obtaining a conclusive outcome~$P^{\rm min}_e(Q)/\Qb$ (dashed-dotted line). For $Q\ge Q_{\rm c}$, the two lines become straight, with a slope of~$\alpha=1-C=1/3$ resp.~$0$.  For the plot~$\theta=\pi/10$.}
     \label{Fig2}
     \end{figure}

Let us now briefly discuss a second example with linearly dependent states, 
which is interesting because such states cannot be unambiguously discriminated. Consider the trine qubit
states
%
$
\ket{\psi_k}=\cos\theta \ket{0}+\mathrm{e}^{\frac{2 i \pi}{3}k} \sin\theta\ket{1}$, $0\le\theta\le\pi/4$, with equal prior probabilities $\eta_k=1/3$,  $ k=1,2,3$. Note that $2\theta$ is the polar angle on the Bloch sphere.
%
%
The set of signal states is covariant with respect to the abelian group of unitaries $\{\id,u,u^2\}$, where $u=|0\rangle\langle0|+\mathrm{e}^{\frac{2 i \pi}{3}}|1\rangle\langle1|$, and
this implies that $\Pi_0$ can be chosen diagonal in the basis~$\{|0\rangle,|1\rangle\}$. Moreover, since an optimal pure strategy requires that $\Pi_0$ has a zero eigenvalue, we must have~$\Pi_0=\xi |0\rangle\langle0|$ (the other possible choice,~\mbox{$\Pi_0=\xi |1\rangle\langle1|$}, turns out  not to be optimal).
Using~Eq.~(\ref{tildestates}) one easily obtains~$|\tilde\psi_k\rangle=\cos\tilde\theta\,|0\rangle+{\rm e}^{{2i\pi\over3}k}\sin\tilde\theta\,|1\rangle$, with $\cos\tilde\theta=\bar\xi{}^{\,1/2} (\bar\xi\cos\!{}^2\theta+\sin\!{}^2\theta)^{-1/2}\!\cos\theta$ and 
$\tilde{\eta}_k=\eta_k=1/3$, i.e., the transformed states are themselves trine states with polar angle $2\tilde\theta$. Eq.~(\ref{Q}) gives~$Q=\xi \cos^2\theta$, where we have used that the averaged density matrix $\rho$ for the trine states is $\rho=(1/3)\sum_{k=1}^3|\psi_k\rangle\langle\psi_k|$, hence $\xi$ is determined, and we simply have $P^{\rm min}_e(Q)=\Qb \tilde P^{\rm ME}_e$, as no minimization over~$\Pi_0$ is possible.  
After some algebra we can rewrite~$\cos\tilde\theta$ as~$\cos\tilde\theta=\Qb{}^{-1/2}(\cos^2\theta-Q)^{1/2}$, and $\sin\tilde\theta=\Qb{}^{-1/2}\sin\theta$. Substituting in $\tilde P^{\rm ME}_e=(2-\sin2\tilde\theta)/3$ (see, e.g.,~\cite{croke}) 
we obtain
\begin{equation}\label{pe-trines}
P^{\rm min}_e(Q)={2\over3}\left(\Qb-\sin\theta \sqrt{\cos^2\theta-Q}\right),\ Q\le Q_{\rm c}.
\end{equation}
%
%
To calculate~$Q_{\rm c}$ we use Eq.~(\ref{alpha}) for $Q=Q_{\rm c}$. For the case at hand it reads $P^{\rm min}(Q_{\rm c})=(Q_{\rm c}-1)(P^{\rm min}_e)'(Q_{\rm c})$, where we have used that $P^{\rm min}_e(Q)$, defined in Eq.~(\ref{pe-trines}), is differentiable.
The solution is~$Q_{\rm c}=\cos 2\theta$, which in turn yields~$\alpha=1/3$. 
Then, $P^{\rm min}_e(Q)$ is given by Eq.~(\ref{alpha}) with these particular values of~$\alpha$, and $Q_{\rm c}$.
%
%
The latter and $\bar\alpha=2/3$ are both in agreement with the values of the optimal failure probability $Q^{\rm MC}$ and the maximum confidence $C$, respectively, for the trine states in~\cite{croke}. These results are illustrated in Fig.~\ref{Fig2}. They exemplify the link between MC and FRIO schemes. 

To summarize, we introduced a very general transformation in Eqs.~(\ref{tildeops}) and~(\ref{tildestates}) that turns every problem with fixed inconclusive rate into an equivalent ME problem. When the solution 
of the resulting ME problem is known one can optimize it over the free parameters of the transformation. In 
some special cases, including symmetric states~\cite{sym,bergourev} or two 
mixed states whose density matrices are diagonal in the Jordan basis \cite{jordan}, this can 
be done analytically. 
%
We have identified a critical value $Q_{\rm c}$ of  inconclusive rate that generalizes the notion of failure probability used in UD to other cases where UD cannot be applied, such as discrimination of linearly dependent states or full rank mixed states. We note that related work has been done independently by Ulrike Herzog \cite{herzog3}. We will present further details in a separate publication \cite{bagan2}. The method we presented here is very powerful and can be applied in many other cases, including quantum state estimation with post processing~\cite{grcm-tb}. 

\begin{acknowledgments}
\noindent\emph{Acknowledgments}. This research was supported by NSF Grant PHY0903660,
the Spanish MICINN, through contract FIS2008-01236,  project QOIT
(CONSOLIDER 2006-00019) and (EB) PR2010-0367, and from the Generalitat de
Catalunya CIRIT, contract  2009SGR-0985. We also acknowledge financial support from ERDF: European Regional Development Fund.
\end{acknowledgments}

\bibliographystyle{unsrt}

\end{document}